\documentclass[12pt]{article}
\usepackage{graphicx}
\usepackage{multirow}
\usepackage{amsmath}
\usepackage{amssymb}
\input{epsf}
\begin{document}
\begin{center}
{\bf APPLICATION OF THE TRIAXIAL QUADRUPOLE-OCTUPOLE \\ROTOR TO THE
GROUND AND NEGATIVE-PARITY\\ LEVELS OF ACTINIDE NUCLEI}
 \vspace*{1cm}

 {\bf M. S. Nadirbekov$^{1,3}$,
 N. Minkov$^2$, M. Strecker$^3$ and W. Scheid$^3$}\\
\vspace*{9 mm}
{\it$^1$Institute of
Nuclear Physics, Ulughbek, Tashkent, 100214, Uzbekistan\\
 $^2$Institute of Nuclear Research and Nuclear
Energy, 72 Tzarigrad Road, BG-1784 Sofia, Bulgaria\\
$^3$Institut $f\ddot{u}r$ Theoretische Physik der
Justus-Liebig-$Universit\ddot{a}t$, Heinrich-Buff-Ring 16, D-35392
Giessen, Germany}
\vspace*{1cm}
\end{center}

\noindent {\bf Abstract:} In this work we examine the possibility to describe yrast positive-
and negative-parity excitations of deformed even-even nuclei through a collective rotation
model in which the nuclear surface is characterized by triaxial quadrupole and octupole
deformations. The nuclear moments of inertia are expressed as sums of quadrupole and
octupole parts. By assuming an adiabatic separation of rotation and vibration degrees of
freedom we suppose that the structure of the positive- and negative- parity bands may be
determined by the triaxial-rigid-rotor motion of the nucleus. By diagonalizing the Hamiltonian
in a symmetrized rotor basis with embedded parity we obtain a model description for the yrast
positive- and negative-parity bands in several actinide nuclei. We show that the energy
displacement between the opposite-parity sequences can be explained as the result of the
quadrupole-octupole triaxiality.

\bigskip


\bigskip\bigskip\bigskip

PACS Number(s): 21.10.Re, 21.60.Ev

\begin{center}
{\bf 1. Introduction}
\end{center}

A basic problem in the collective excited states of even-even nuclei is the connection
bet\-ween the rotational motion and surface deformations with different multipolarities
\cite{EG87}.  It is considered that the even-multipolarity deformations give rise to
positive-parity states, while the odd-multipolarities can lead to the appearance of
negative-parity states. Also, it is considered that the axial deformations play the leading role in
the nuclear rotation collectivity, whereas the non-axial ones may be responsible for some
specific properties of the spectra. Collective excited states with positive and/or negative
parities in the cases of axial and non-axial deformations are explained by the early models
\cite{Bo52}--\cite{6}. In \cite{4} an adiabatic approximation (separation of rotation motion
from vibrations) was applied, providing a possibility for description of positive-parity spectra
of even-even nuclei for which the presence of non-axial quadrupole deformation is assumed.
Also, some specific properties of rotational bands can be associated with the presence of
effective quadrupole and octupole non-axial deformations. Based on such an assumption, the
model approaches \cite{5}-\cite{19} attempt the simultaneous description of both, positive
and negative-parity states, in even-even nuclei. From another side, it is commonly accepted
that the appearance of negative-parity states is mainly determined by the octupole
(reflection-asymmetric) deformations with the main contribution of the axial deformation
mode \cite{EG87,BN96}. The spectroscopic properties of nuclei with assumed axial
quadrupole and octupole deformations are considered in detail in
\cite{Krappe69}--\cite{NYMS12} for different types of the potential energy depending on the
surface deformation.

In this work a collective rotation model with non-axial quadrupole and octupole deformations
is considered under the adiabatic approximation. It is applied for the simultaneous description
of the ground-state (g.s.) and the lowest negative-parity bands of the even-even nuclei
$^{228-232}$Th, $^{230-238}$U and $^{240}$Pu. In Refs. \cite{5,17,18,19} the non-axial
deformations are treated similarly to the model of Davydov-Chaban \cite{20}, but without the
use of the adiabatic approximation. In the present work the advantages of the adiabatic
approximation are used to formulate a simplified pure rotor problem with respect to the
nuclear surface parametrization explored in \cite{19}. As a result the g.s. (positive-parity) and
negative-parity levels of deformed even-even nuclei are considered as the excitations of a
triaxial quadrupole-octupole rotor. In this approach the energy shift between the
opposite-parity states is obtained as the result of the $K$-mixing effect due to the presence of
quadrupole and octupole non-axial deformations. Based on the earlier interpretation \cite{19}
of the positive- and negative-parity excitations, the present approach is essentially different
from the concept of pure octupole, or mixed quadrupole-octupole, vibrations and rotations.
The latter concept assumes nuclear vibrations with respect to an octupole double-well
\cite{Krappe69}--\cite{Jolos94}, \cite{MYDS06}, or more general two-dimensional
quadrupole-octupole  potential  with axial symmetry \cite{10}--\cite{16},
\cite{14}--\cite{NYMS12} and rotations, used to explain the so-called alternating-parity
spectra.

The purpose of this work is to quantitatively examine the possibility to explain the structure of
alternating-parity bands within the triaxial quadrupole-octupole rotor concept based on the
nowadays experimental data \cite{21} on positive- and negative-parity energy levels in heavy
even-even nuclei. As it will be seen below, the result can be compared to descriptions
obtained within the quadrupole-octupole vibration-rotation concept. We think that such an
investigation would provide a useful framework to evaluate the separate roles of the axial and
non-axial degrees of freedom in the forming of collective spectra in the nuclei with
quadrupole and octupole deformations as well as to examine the possibility to combine them
into a common model approach.

In sec. 2 the quadrupole--octupole parametrization of the nuclear surface is briefly recalled. In
sec. 3  the triaxial quadrupole--octupole rotor Hamiltonian and the solution of the eigenvalue
problem for the alternating-parity excitations are presented. In sec. 4 results of the numerical
application of the model to several even-even actinide nuclei are presented together with a
relevant analysis and discussion of the model dynamic mechanism. In sec. 5 concluding
remarks are given.

\begin{center}
{\bf 2. Quadrupole--octupole parametrization of the nuclear surface}
\end{center}

The distance between the center of the deformed nucleus and its surface in the direction of the
polar angles $\theta$, $\varphi$ in the laboratory frame $x$, $y$, $z$ is given by the
expression \cite{EG87,1,2,3,4}:
\begin{equation}
R(\theta,\varphi)=R_0\left[1+\sum_{\lambda\mu}\alpha_{\lambda\mu}
Y^{*}_{\lambda\mu}(\theta,\varphi)\right],
\end{equation}
where $R_0$ is the radius of the spherical shape. The parameters of nuclear deformations
$\alpha_{\lambda\mu}$ satisfy the condition
$\alpha^*_{\lambda\mu}=(-1)^\mu\alpha_{\lambda,-\mu}$, and for the spherical harmonics
one has $Y^{*}_{\lambda\mu}(\theta,\varphi)=
(-1)^{\mu}Y_{\lambda,-\mu}(\theta,\varphi)$. In the case of quadrupole and octupole
deformations expression (1) can be written as
\begin{equation}
R(\theta,\varphi)=R_0\left[1+\sum_{\mu=-2}^{2}
\alpha_{2\mu}Y^{*}_{2\mu}(\theta,\varphi)+
\sum_{m=-3}^{3}\alpha_{3m}Y^{*}_{3m}(\theta,\varphi)\right].
\end{equation}

We introduce an intrinsic coordinate system the origin of which is fixed at the center of mass,
with the axes $\xi$, $\eta$, $\zeta$ being directed along the principal axes of inertia of the
nucleus. The orientation of the intrinsic axes with respect to the axes $x$, $y$, $z$ is defined
by the Euler angles $\theta=\{\theta_1, \theta_2, \theta_3\}$, such that
\begin{equation}
R(\theta',\varphi')=R_0\left[1+\sum_{\nu=-2}^{2}
a_{2\nu}Y^{*}_{2\nu}(\theta',\varphi')+
\sum_{m=-3}^{3}a_{3m}Y^{*}_{3m}(\theta',\varphi')\right],
\label{Rintr}
\end{equation}
with
$$a_{\lambda\nu}=\sum_{\mu}\alpha_{\lambda\mu}D^{\ast\lambda}_{\nu\mu}(\theta),
\hspace{1cm}
\alpha_{\lambda\mu}=\sum_{\nu}D^{\lambda}_{\mu\nu}(\theta)a_{\lambda\nu},
\qquad \lambda=2,3,$$ where $D^{\lambda}_{\mu\nu}(\theta)$ is the
Wigner function according to \cite{wigner}.

A possible parametrization  \cite{1,17} of the quadrupole-octupole shape in the intrinsic coordinate
system is
\begin{equation}
a_{2,1}=a_{2,-1}=0, \hspace{1cm} a_{3,\pm1}=a_{3,\pm3}=0
\label{azero}
\end{equation}
and
\begin{equation}
a_{2,0}=\beta_2\cos\gamma, \hspace{1cm}
a_{2,2}=a_{2,-2}=\frac{\beta_2\sin\gamma}{\sqrt{2}},
\label{agamma}
\end{equation}
\begin{equation}
a_{3,0}=\beta_3\cos\eta, \hspace{1cm} a_{3,2}=a_{3,-2}=\frac{\beta_3\sin\eta}{\sqrt{2}},
\label{aeta}
\end{equation}
where $\beta_2\geq0$ and $\gamma$ ($0\leq\gamma\leq\frac{\pi}{3}$ in the case of pure
quadrupole deformation) are the parameters of axial and triaxial quadrupole deformations,
respectively \cite{1}; $\beta_3$ and $\eta$ ($0\leq\eta\leq\frac{\pi}{2}$ in the case of pure
octupole deformation) are the parameters of axial and triaxial octupole deformations,
respectively \cite{17}.  We remark that the case $a_{3,\pm1}\neq0, a_{3,\pm3}\neq0$ was
considered in \cite{18}, whereas in \cite{solov} it was shown that deformation degrees of
freedom determined by $a_{3,\pm3}$ can be associated with pure two-quasiparticle states
and may not need to be considered as collective variables. Also, in the above parametrization
(\ref{azero})-(\ref{aeta}) the octupole shape is aligned with respect to the quadrupole
ellipsoid which determines the body-fixed frame.

\begin{center}
{\bf 3. Triaxial quadrupole-octupole rotor in adiabatic approximation}
\end{center}

After imposing the adiabatic approximation the collective rotation motion of
the nucleus is considered separately from the other degrees of freedom. If no
further assumptions (e.g. axial symmetry) are imposed, the corresponding
energy spectrum is described by the triaxial-rotor Hamiltonian
\begin{equation}
\hat{H}_{\mbox{\scriptsize rot}}=\sum^3_{i=1}
\frac{\hbar^2\hat{I}^2_i}{2\mathfrak{J}_i},
\label{Hrot}
\end{equation}
where $\hat{I}_i$ ($i=1,2,3$) are the components of the angular momentum
operator and $\mathfrak{J}_{i}$ are the respective components of the moment
of inertia.  By applying the parametrization (\ref{azero})-(\ref{aeta}) into
the expansion (\ref{Rintr}) the latter are obtained in a form corresponding
to a triaxial quadrupole-octupole shape \cite{19}
\begin{eqnarray}
\mathfrak{J}_1&=&8B_2\beta_2^2\sin^2\left(\gamma-\frac{2\pi}{3}\right)
+8B_3\beta_3^2\left[\frac{3}{2}\cos^2\eta+\sin^2\eta+\frac{\sqrt{15}}{2}
\sin\eta \cos\eta\right]\label{J1}\\
\mathfrak{J}_2&=&8B_2\beta_2^2\sin^2\left(\gamma-\frac{4\pi}{3}\right)
+8B_3\beta_3^2\left[\frac{3}{2}\cos^2\eta+\sin^2\eta-\frac{\sqrt{15}}{2}
\sin\eta \cos\eta\right]\label{J2}\\
\mathfrak{J}_3&=&8B_2\beta_2^2\sin^2\gamma+8B_3\beta_3^2\sin^2\eta
\label{J3},
\end{eqnarray}
where $\mathfrak{J}_i$ ($i=1,2,3$) explicitly depend on the quadrupole and octupole mass
and deformation parameters $B_2$, $\beta_2$, $\gamma$ and $B_3$, $\beta_3$, $\eta$,
respectively.

In the present work we consider (under the assumption of the adiabatic approximation) the
deformation parameters (or more precisely their average values)  as constants whose values
$\beta_{2\, \mbox{\scriptsize eff}}$, $\gamma_{\mbox{\scriptsize eff}}$ and $\beta_{3\,
\mbox{\scriptsize eff}}$, $\eta_{\mbox{\scriptsize eff}}$, effectively determine a rigid triaxial
rotor. Since the mass parameters and the squares of the axial deformation parameters enter the
moments of inertia components as \emph{products} we introduce in (\ref{J1})--(\ref{J3}) the
following parameters $\tilde{B}_2$=$8B_2\beta_{2\, \mbox{\scriptsize eff}}^{2}$ and
$\tilde{B}_3$=$8B_3\beta_{3\, \mbox{\scriptsize eff}}^{2}$. Thus we obtain the
quadrupole-octupole moment of inertia in a form depending on \emph{four} arguments
$\mathfrak{J}_i=\mathfrak{J}_i(\tilde{B}_2, \tilde{B}_3, \gamma_{\mbox{\scriptsize
eff}},\eta_{\mbox{\scriptsize eff}})$. We take these arguments, $\tilde{B}_2, \tilde{B}_3,
\gamma_{\mbox{\scriptsize eff}},\eta_{\mbox{\scriptsize eff}}$, as \emph{adjustable
parameters}. By introducing the reciprocal inertia factors $A_i(\tilde{B}_2, \tilde{B}_3,
\gamma_{\mbox{\scriptsize eff}},\eta_{\mbox{\scriptsize
eff}})=1/[2\mathfrak{J}_i(\tilde{B}_2, \tilde{B}_3, \gamma_{\mbox{\scriptsize
eff}},\eta_{\mbox{\scriptsize eff}})]$ ($i=1,2,3$) we get the Hamiltonian (\ref{Hrot}) in the
form
\begin{eqnarray}
\hat{H}_{\mbox{\scriptsize rot}}=\sum^3_{i=1} \hbar^2A_{i}(\tilde{B}_2, \tilde{B}_3,
\gamma_{\mbox{\scriptsize eff}},\eta_{\mbox{\scriptsize
eff}})\hat{I}^2_{i}
\end{eqnarray}
depending on the four parameters $\tilde{B}_2, \tilde{B}_3,
\gamma_{\mbox{\scriptsize eff}},\eta_{\mbox{\scriptsize eff}}$.

Here it is important to remark that the involvement of the quantities
$\tilde{B}_2$ and $\tilde{B}_3$ assumes a more general dependence of the
moment of inertia on the respective mass and axial-deformation parameters. In
this way the solution of the rotation energy problem does not constrain the
mass and deformation parameters separately but only determines their
products. In the context of the more general rotation-vibration problem (not
considered here) this means that $\tilde{B}_2$ and $\tilde{B}_3$ can be
interpreted as generalized mass parameters carrying implicit dependence (not
necessarily quadratic) on the axial quadrupole and octupole deformations.
From this aspect the situation resembles the concept of
deformation-dependent-mass approaches where some plausible dependencies of
the mass parameter on the axial quadrupole deformation are assumed under
proper physical argumentation \cite{Jolos08,Jolos09,DDM10,DDM11,DDM13}. As
here the vibration energy is not considered, the parameters $\tilde{B}_2$ and
$\tilde{B}_3$ only determine the general scale of the rotation energy and the
relative contributions of the quadrupole and octupole deformations in the
moment of inertia, while their particular deformation dependence does not
play a role.

The energy spectrum of the above determined triaxial quadrupole-octupole
rotor is obtained by diagonalizing the Hamiltonian (\ref{Hrot}) in the basis
of the symmetrized rotor functions with built-in parity \cite{19}
\begin{eqnarray}
|IMK \pm\rangle=\frac{1}{\sqrt{2(1+\delta_{K,0})}}
\left(|IMK\rangle\pm (-1)^{I-K}|IM-K\rangle \right) \ ,
\label{basis}
\end{eqnarray}
where $|IMK\rangle =\sqrt{\frac{2I+1}{8\pi^2}}D^{I}_{MK}(\theta)$, and $M$
and $K$ are the projections of the angular momentum $\hat{I}$ on the third
axes of the laboratory and intrinsic frames, respectively. The eigenfunctions
are obtained in the form
\begin{equation}
\Phi_{IMn}^{\pm}(\theta)=\sum_{K\geq
0}^{I}C^n_{IK}|IMK\pm\rangle \ .
\label{wfunc}
\end{equation}
The expansion coefficients implicitly depend on the effective mass and
deformation parameters: $C^n_{IK}=C^n_{IK}(\tilde{B}_2, \tilde{B}_3,
\gamma_{\mbox{\scriptsize eff}},\eta_{\mbox{\scriptsize eff}})$. For given
angular momentum $I$ the quantum number $n = 0, 1, 2, \dots$ (limited by
$K\leq I$) labels the different eigenvalues $E^n(I)$ of
$\hat{H}_{\mbox{\scriptsize rot}}$ in ascending order.

We obtain the matrix elements of Hamiltonian (\ref{Hrot}) between basis states (\ref{basis})
with different $K$-values in the following general form
\begin{eqnarray}
& &\langle IMK'\pm |\hat{H}_{\mbox{\scriptsize rot}}|IMK\pm\rangle
=\frac{1}{\sqrt{(1+\delta_{K',0})(1+\delta_{K,0})}}\nonumber \\
&\times&\left \{ \frac{1}{2}\left[(A_{1}+A_{2})I(I+1)-(A_{1}+A_{2}-2A_{3})K^{2}
\right] \left[\delta_{K'K}\pm (-1)^{I-K}\delta_{K',-K}\right]\right.\nonumber \\
&+& (A_{1}-A_{2})f(I,K)\left[ \delta_{K',K+2}\pm
(-1)^{I-K}\delta_{K',-K-2}\right]\nonumber \\
&+&(A_{1}-A_{2})f(I,-K)\left[ \delta_{K',K-2}\pm
(-1)^{I-K}\delta_{K',-K+2}\right] \biggr\}, \label{matel}
\end{eqnarray}
with
\begin{eqnarray}
f(I,K)=\frac{1}{4}\sqrt{(I+K+2)(I+K+1)(I-K-1)(I-K)}.
\end{eqnarray}
Since $K,K'\geq 0$, in (\ref{matel}) one has $\delta_{K',-K}\neq 0$ only for
$K=0$, $\delta_{K',-K+2}\neq 0$ only for $K=0,1,2$, while $\delta_{K',-K-2}$ is
always zero. As a result Eq. (\ref{matel}) can be written in a bit simpler form
\begin{eqnarray}
& &\langle IMK'\pm |\hat{H}_{\mbox{\scriptsize rot}}|IMK \pm\rangle
=\frac{1}{2\sqrt{(1+\delta_{K',0})(1+\delta_{K,0})}}\nonumber \\
&\times&\left \{ \left[(A_{1}+A_{2})I(I+1)-(A_{1}+A_{2}-2A_{3})K^{2} \right]
\left[\delta_{K'K}\pm (-1)^{I-K}\delta_{K',-K}\right]\right.\nonumber \\
&+& \left. 2(A_{1}-A_{2})\left[f(I,K)\delta_{K',K+2}+ f(I,-K)\left(
\delta_{K',K-2}\pm (-1)^{I-K}\delta_{K',-K+2}\right)\right] \right \}.
\label{matel1}
\end{eqnarray}

Here we make the following important assumptions originally imposed in
\cite{19}:
\smallskip

(i) The quantum number $K$ takes only even values (see also \cite{5,2,6}). Then in (\ref{wfunc}) the sum over
$K$ runs from 0 to $I$ in steps of $2$.
\smallskip

(ii) The $(+)$ and $(-)$ phases in (\ref{wfunc}) and(\ref{basis}), which correspond to the
$A$ and $B_{1}$ classes of the rotation group $D_{2}$, are attributed to the positive- and
negative-parity states, respectively.
\smallskip

Because of (i) the phase factor $(-1)^{K}=1$ can be omitted in the matrix element expressions
(\ref{matel}) and (\ref{matel1}). According to (ii) one can write $ \Phi_{IMn}^{\pm} \equiv
\Phi_{IMn}^{\pi}$ and $|IMK \pm\rangle \equiv |IMK \pi\rangle$ in (\ref{wfunc}) and
(\ref{basis}), respectively, where $\pi =\pm$ is the parity of the rotation state. Also, according
to (\ref{basis}) the $\pi =(+)$ parity corresponds to $I=$ even angular momentum states and
$\pi =(-)$ corresponds to $I=$ odd, as in the usual alternating-parity (octupole) bands.
Therefore, one has $\pi =(-1)^{I}$. Then the $(\pm)$ and phase factors in (\ref{matel}) and
(\ref{matel1}) can be omitted when alternating-parity bands are considered. As a result the
matrix element (\ref{matel1}) reads
\begin{eqnarray}
& &\langle IMK'\pi |\hat{H}_{\mbox{\scriptsize rot}}|IMK\pi\rangle
=\frac{1}{2\sqrt{(1+\delta_{K',0})(1+\delta_{K,0})}}\nonumber \\
&\times&\left \{ \left[(A_{1}+A_{2})I(I+1)-(A_{1}+A_{2}-2A_{3})K^{2} \right]
\left[\delta_{K'K}+\delta_{K',-K}\right]\right.\nonumber \\
&+& \left. 2(A_{1}-A_{2})\left[f(I,K)\delta_{K',K+2}+ f(I,-K)\left(
\delta_{K',K-2}+\delta_{K',-K+2}\right)\right] \right \}.
\label{matel2}
\end{eqnarray}
Note that the righthand side of (\ref{matel2}) does not include $\pi$ explicitly, whereas as
mentioned above, it is implied by the value of $I$ (even or odd). The lowest eigenvalues $E^{n =0}(I)$ and
eigenfunctions $\Phi_{IM(n =0)}^{\pi}$ of the Hamiltonian correspond to the g.s. band for
$I=$even, and to the lowest negative-parity band for $I=$odd. In the same way the higher
eigenfunctions/eigenvalues with $n =1,2, ...$ determine higher sets of positive- and
negative-parity bands. In this work we only consider the lowest (yrast) band structure with $n
=0$.

Now, one can easily check that due to the non-diagonal matrix elements in (\ref{matel2}),
which mix the basis states with $\Delta K =2$, the negative-parity levels appear shifted up in
energy with respect to the positive-parity levels in the g.s. band. The magnitude of the shift
depends on the values of the non-diagonal matrix elements. It increases with the increase of
the quadrupole and octupole non-axiality parameters $\gamma_{\mbox{\scriptsize eff}}$ and
$\eta_{\mbox{\scriptsize eff}}$ and decreases with the increase of the angular momentum. In
this way the considered  triaxial quadrupole-octupole rotor approach allows a possibility to
reproduce the observed behaviour of the lowest nuclear positive- and negative-parity bands,
which are often interpreted as an yrast alternating-parity band.

In the present approach the staggering effect observed in the
alternating-parity bands appears as the result of two model features: 1)
association of the parity of the observed rotation states (the +/- phases in
(\ref{wfunc}) and (\ref{basis})) with the parity of the angular momentum
values, and 2) mixing of different K-modes due to the assumed triaxial
deformations.

We remark that although the triaxial rotor Hamiltonian (\ref{Hrot}) mixes
different K-modes it does not mix the parity. Therefore, the
rotation states appearing in the model have a well-defined parity.
It should be noted that also the intrinsic state which is presently
adiabatically separated is usually considered with a good parity.
Thus, when it is associated with a double-well potential one gets
the lowest potential level whose (determined) parity combined with
the parity of the rotation-mode provides a good parity of the total
(observed) state. More details on that issue are given e.g. in ref.
\cite{MYDS06}. In the present approach, due to the adiabatic approximation,
the intrinsic parity is directly built in the parity of the
observed rotation state through the parity of the angular momentum
value $(-1)^I$.

\begin{center}
{\bf 4. Numerical results and discussion}
\end{center}

The theoretical spectrum obtained through the above diagonalization procedure
was applied to \emph{simultaneously} describe the yrast positive- and
negative-parity levels of several actinide nuclei: $^{228,230,232}$Th,
$^{230,232,234,236,238}$U and $^{240}$Pu and shown in Figs. 1-9. The triaxial
quadrupole-octupole rotor parameters $\tilde{B}_2$, $\tilde{B}_3$,
$\gamma_{\mbox{\scriptsize eff}}$ and $\eta_{\mbox{\scriptsize eff}}$ were
adjusted to experimental data through a least-square fitting procedure. The
theoretical and experimental energy values are compared in Figs.~1-9 (left).
To get more detailed information about the angular momentum behaviour of the
theoretical and experimental alternating-parity sequences we apply the
following  odd-even staggering formula \cite{14}
\begin{equation}
Stg(I)=6\Delta E(I)-4\Delta E(I-1)-4\Delta E(I+1)+\Delta
E(I+2)+\Delta E(I-2),
\end{equation}
with $\Delta E(I)=E(I+1)-E(I)$. The obtained theoretical and experimental staggering patterns
are presented in Figs.~1-9 (right). The adjusted parameter values and the root-mean-square
(RMS) deviations between theoretical and experimental energy values are given in the
captions.

Here, we can make the following comments.

In the nuclei  $^{228}$Th (RMS=66.2 keV),  $^{230}$U (RMS=70.3 keV), $^{232}$U
(RMS=39.4 keV), $^{234}$U (RMS=11.3 keV) and $^{238}$U (RMS=68.3 keV) the model
description is relatively good. The obtained RMS values are comparable with the values
obtained in models where the alternating-parity bands are described as the result of mixed
quadrupole-octupole vibrations and rotations \cite{14,MDSSL12,NYMS12}. We remark that
in the nuclei $^{230,232,234, 238}$U the staggering amplitude  shows a slow decrease without
reaching zero.

In the remaining nuclei,   $^{230}$Th (RMS=119.3 keV), $^{232}$Th (RMS=160 keV),
$^{236}$U (RMS= 145.5 keV) and $^{240}$Pu (RMS=132.4 keV)  the deviation between
theory and experiment is essentially larger with RMS$>$100 keV.  This is also observed in the
staggering diagrams in figures 2, 3, 7 and 9, where the experimental staggering amplitude
decreases more rapidly compared to the theoretical one. It is clear that in these cases the
$K$-mixing effect generated by the model triaxial rotations remains too strong in the high
angular momentum $I$. As a result the energy shift between positive- and negative-parity
sequences in the theoretical spectrum remains large, causing a more persistent staggering
effect. Thus, the model reproduces with a less accuracy the fine structure of spectra where a
single (less perturbed) alternating-parity band is formed at high angular momentum. We
remark that in three of the nuclei, $^{228,230}$Th and $^{240}$Pu, the experimental
staggering reaches zero amplitude at high angular momentum and then reappears with a
changed phase. This behaviour is known as a ``beat'' staggering effect, and corresponds to
more complicated properties of the rotating quadrupole-octupole nucleus. It was explained,
e.g. in the so-called quadrupole-octupole rotation model (QORM) \cite{MYDS06}, and
remains beyond the present approach.

Also, we should remark that a more general source of discrepancy between theory and
experiment in all nuclei under study is the fact that we consider a rigid rotor which is, of
course, a rather robust assumption. We see that in the nuclei $^{230,232,234, 238}$U, where the
experimental spectrum is developed up to a not very high angular momentum, the model
effect of rigidity is small, allowing an overall good description of the experimental bands.

To get more detailed insight into the physical contents of the results obtained we have to
analyze the model parameters values. First we note that the effective mass contribution,
$\tilde{B}_2\sim 100$ $\hbar^{2}MeV^{-1}$, of the quadrupole deformation is by two orders
of magnitude larger than the octupole one $\tilde{B}_3\sim 1$ $\hbar^{2}MeV^{-1}$. This
means that the quadrupole deformation enters the problem as a leading collective mode,
whereas the octupole deformation is superposed as a small correction. We
notice that $\tilde{B}_2$ and $\tilde{B}_3$ values used in the model are very narrow
localized around 100 $\hbar^{2}MeV^{-1}$ and 1 $\hbar^{2}MeV^{-1}$, respectively, which
means that the overall energy scale of the collective motion of the different nuclei considered
in the model is quite uniquely determined.

Also, we see that the obtained values of the triaxiality parameters vary in quite narrow limits,
with the quadrupole-triaxiality $\gamma_{\mbox{\scriptsize eff}}$ varying between
50$^{\circ}$ and 57$^{\circ}$, and the octupole-triaxiality $\eta_{\mbox{\scriptsize eff}}$
varying between 48$^{\circ}$ and 50$^{\circ}$.

The obtained results suggest:

i) in the considered nuclei the energy shift between even and odd angular momentum levels
with positive and negative parity, respectively, can be generated by the simultaneous presence
of quadrupole and octupole triaxial deformations. As such it might be associated with the
$K$-mixing and not only with the parity effect.

ii) The $\gamma$-deformation approaching 60$^{\circ}$ corresponds to an overall oblate
quadrupole shape, whereas the presence of the octupole-deformation components suggests a
more complicated shape structure. Although the present model descriptions contradict the
common understanding about pronounced axial deformations manifesting in the heavy
actinide nuclei, this study shows that the triaxial rotor concept may serve as a basis for an
extended consideration of nuclear quadrupole-octupole rotations and vibrations with the
presence of non-axial degrees of freedom.

\begin{center}
{\bf 5. Conclusion}
\end{center}

In this work we evaluated the capability of the triaxial quadrupole-octupole rotor to describe
the structure of the lowest positive- and negative-parity levels in the spectra of heavy
even-even nuclei. It is shown that by using two (quadrupole and octupole) mass parameters
and two (quadrupole and octupole) triaxiality parameters, all of them varying in the
considered nuclei in quite narrow limits. We are able to reproduce the overall structure of the
lowest alternating-parity energy sequences in several actinide nuclei. Although the result is
comparable to the descriptions obtained within the quadrupole-octupole vibration-rotation
concept, it corresponds to a different collective mechanism based on the $K$-mixing effect.
The latter is due to the assumption of strong quadrupole and octupole triaxialities (with the
quadrupole deformation being the leading mode) which according to the model are
responsible for the energy shift between the positive- and negative-parity sequences. On this
basis we conclude that the triaxial quadrupole-octupole rotor approach could provide a useful
tool to evaluate and compare the contributions of the axial and non-axial quadrupole and
octupole deformations in the collective motion of nuclei. Further more detailed
investigation beyond the adiabatic approximation, including the related vibration modes, will
be necessary to connect the different approaches to the problem of alternating-parity spectra
in even-even nuclei.

\begin{center}
{\bf Acknowledgments}
\end{center}

Dr. M. S. Nadirbekov thanks Prof. W. Scheid for the invitation under the DAAD program to
the Institute for Theoretical Physics (Giessen, Germany). Financial supports from the
Uzbekistan Academy of Sciences under contract No. F2-FA-F117 and by the Bulgarian
National Science Fund under contract No. DFNI-E02/6 are gratefully acknowledged.

\newpage

\begin{figure}
\includegraphics[width=0.5\textwidth]{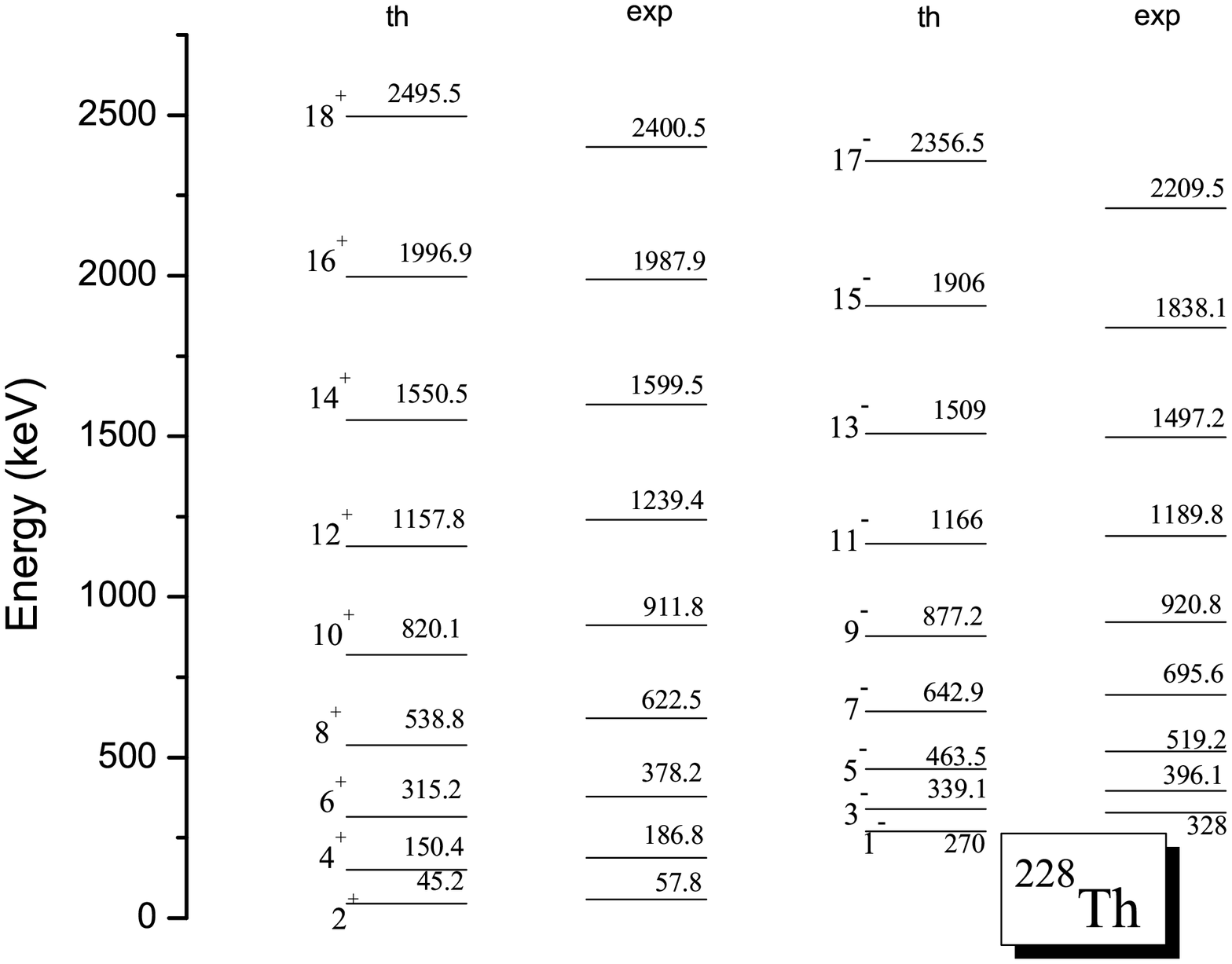}%
\includegraphics[width=0.5\textwidth]{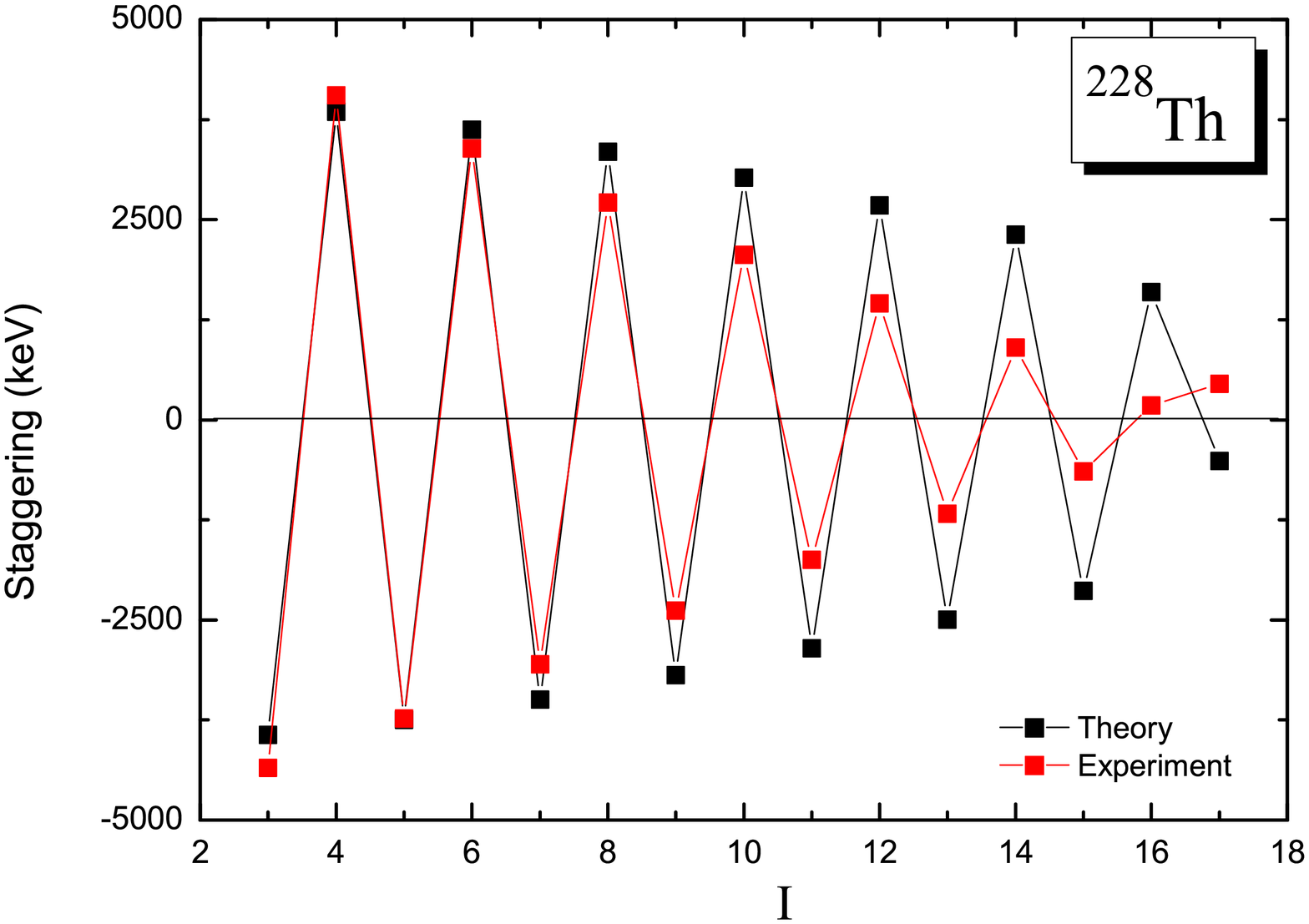}
\caption{Theoretical and experimental energy levels (left) and odd-even staggering patterns (right) for the nucleus
$^{228}$Th (parameters values: $\tilde{B}_2$=90.2$\hbar^{2}MeV^{-1}$,
$\tilde{B}_3$=1$\hbar^{2}MeV^{-1}$, $\gamma_{\mbox{\scriptsize eff}}$=52.2586$^{\circ}$,
$\eta_{\mbox{\scriptsize eff}}$=48.3071$^{\circ}$ and RMS=66.2 keV).}
\end{figure}

\begin{figure}
\includegraphics[width=0.5\textwidth]{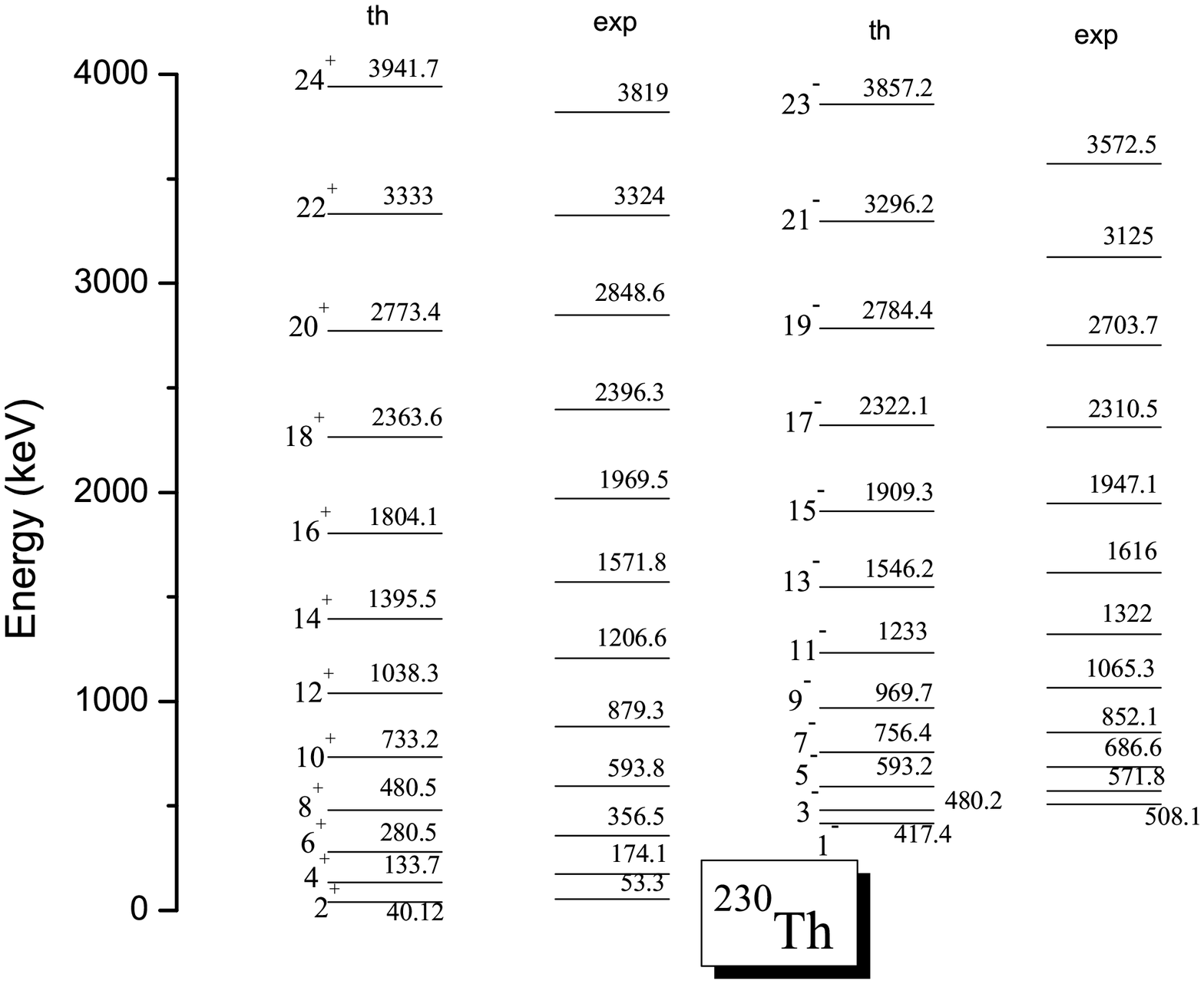}%
\includegraphics[width=0.5\textwidth]{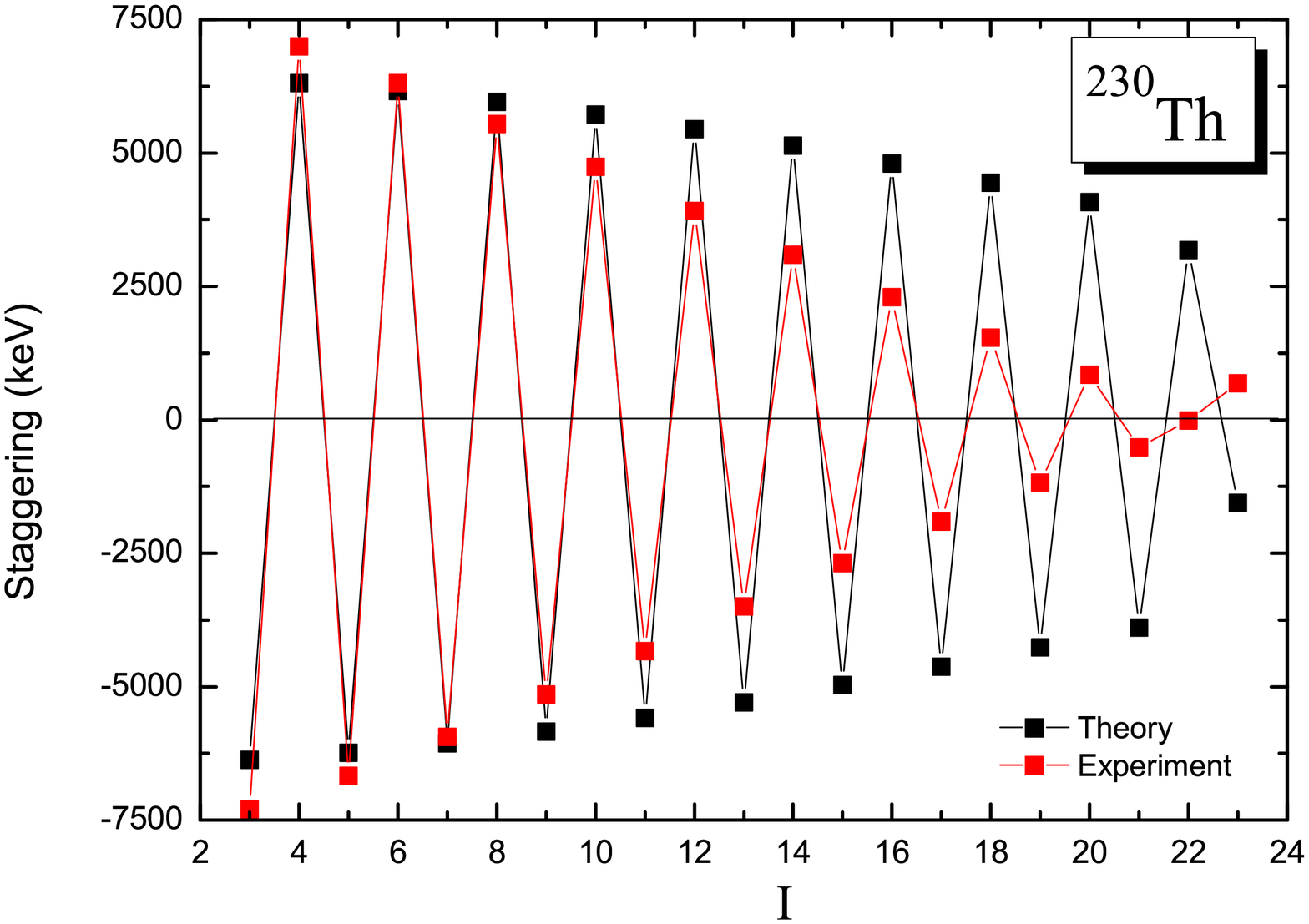}
\caption{The same as in Fig.1, but for
$^{230}$Th (parameters values: $\tilde{B}_2$=100$\hbar^{2}MeV^{-1}$,
$\tilde{B}_3$=1$\hbar^{2}MeV^{-1}$, $\gamma_{\mbox{\scriptsize eff}}$=54.3776$^{\circ}$,
$\eta_{\mbox{\scriptsize eff}}$=49.3124$^{\circ}$ and RMS=119.3 keV).}
\end{figure}

\begin{figure}
\includegraphics[width=0.5\textwidth]{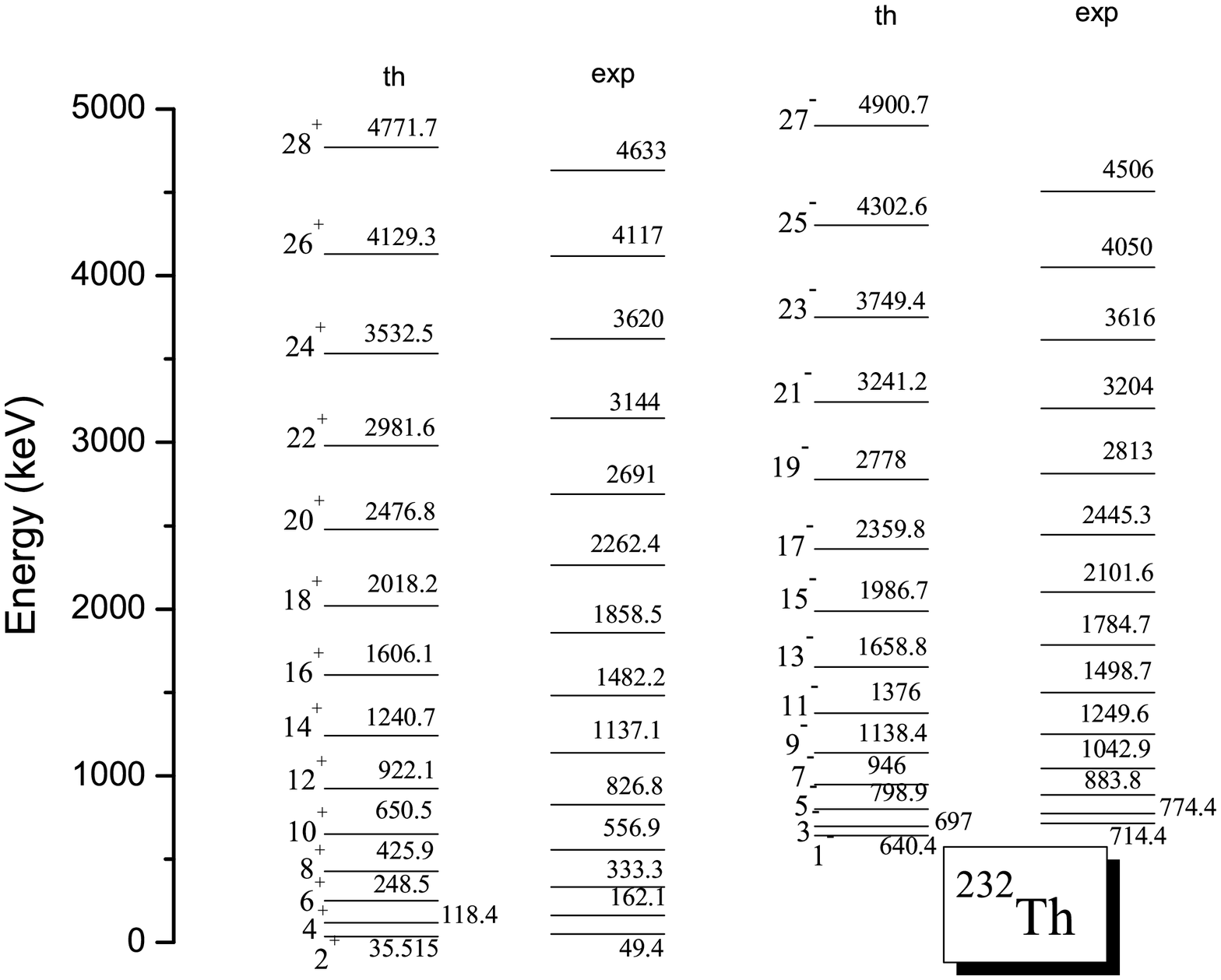}%
\includegraphics[width=0.5\textwidth]{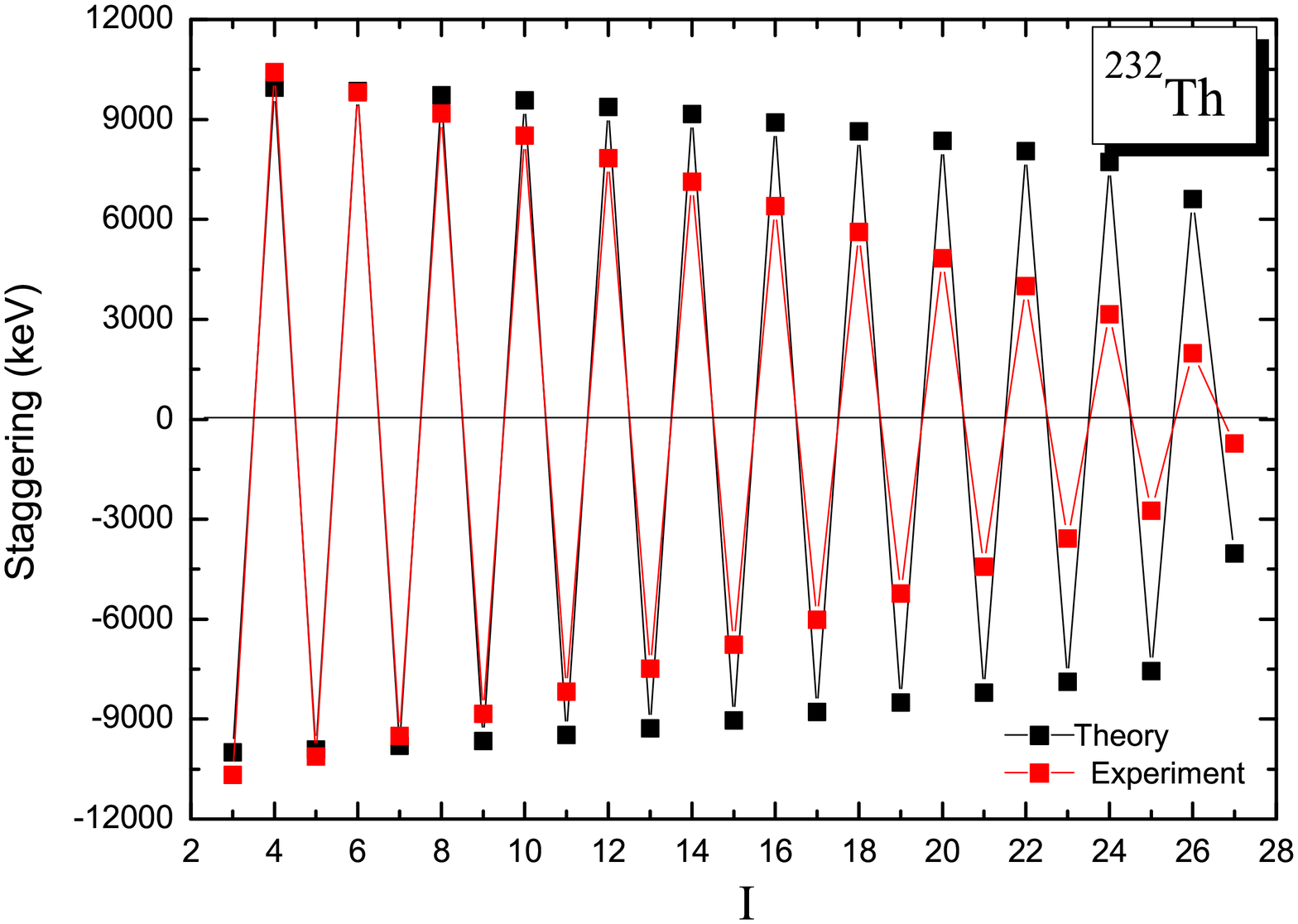}
\caption{The same as in Fig.1, but for  $^{232}$Th (parameters values:
 $\tilde{B}_2$=112$\hbar^{2}MeV^{-1}$,
$\tilde{B}_3$=1.011$\hbar^{2}MeV^{-1}$, $\gamma_{\mbox{\scriptsize eff}}$=56.0482$^{\circ}$,
$\eta_{\mbox{\scriptsize eff}}$=50.1441$^{\circ}$ and RMS=160 keV).}
\end{figure}

\begin{figure}
\includegraphics[width=0.5\textwidth]{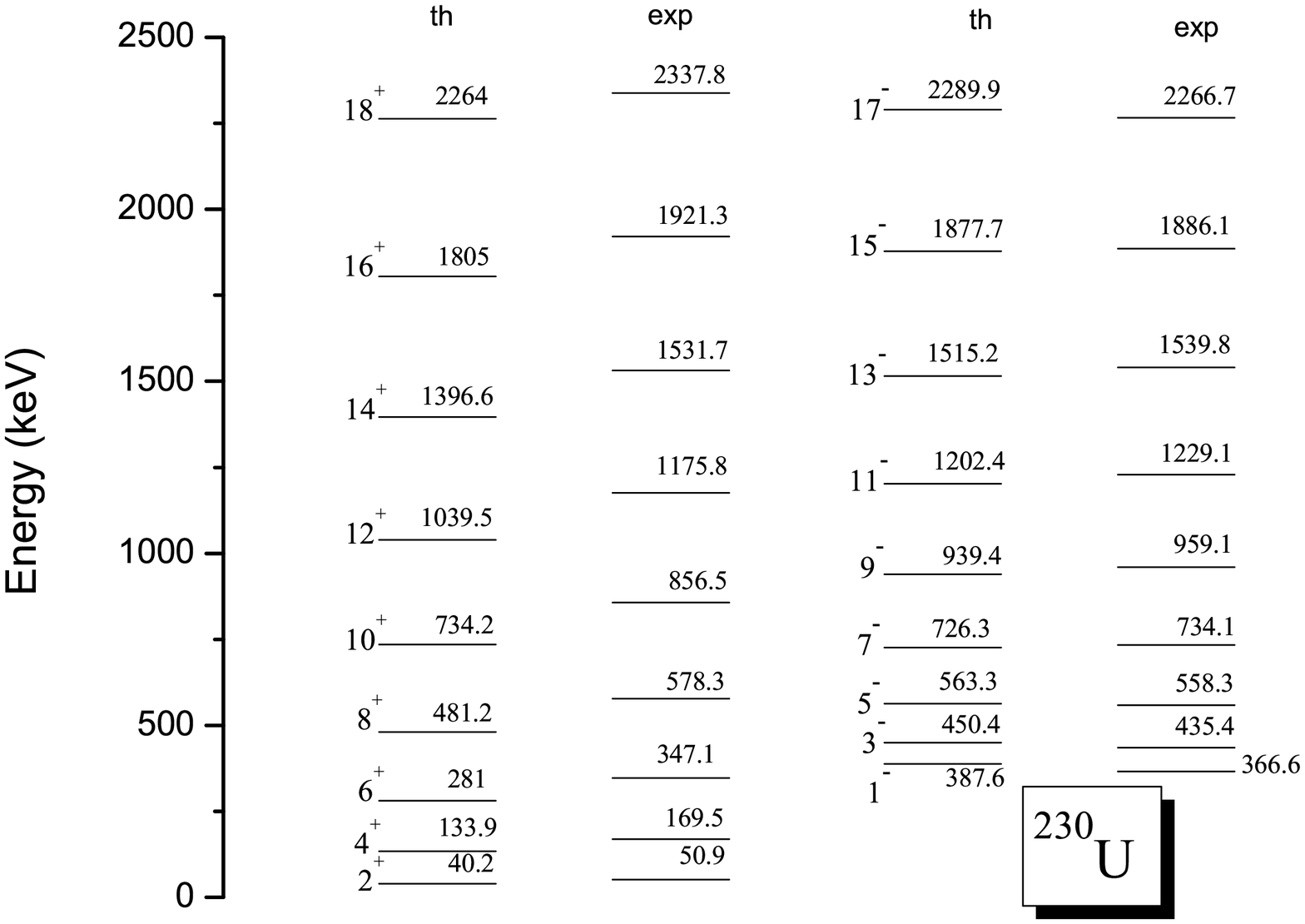}%
\includegraphics[width=0.5\textwidth]{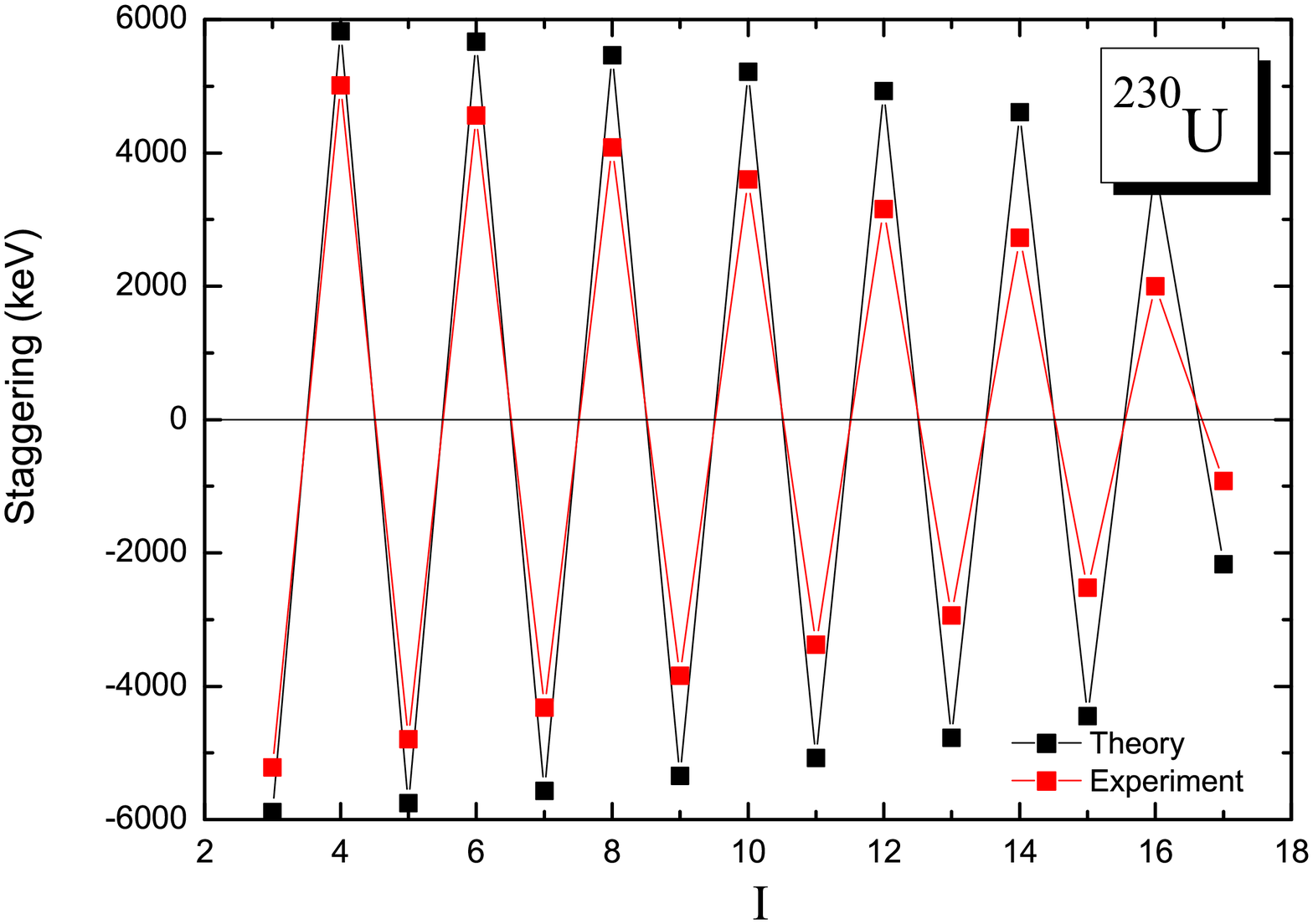}
\caption{The same as in Fig.1, but for $^{230}$U (parameters values:
 $\tilde{B}_2$=100$\hbar^{2}MeV^{-1}$,
$\tilde{B}_3$=1$\hbar^{2}MeV^{-1}$, $\gamma_{\mbox{\scriptsize eff}}$=54.1076$^{\circ}$,
$\eta_{\mbox{\scriptsize eff}}$=49.1587$^{\circ}$ and RMS=70.3 keV).}
\end{figure}

\begin{figure}
\includegraphics[width=0.5\textwidth]{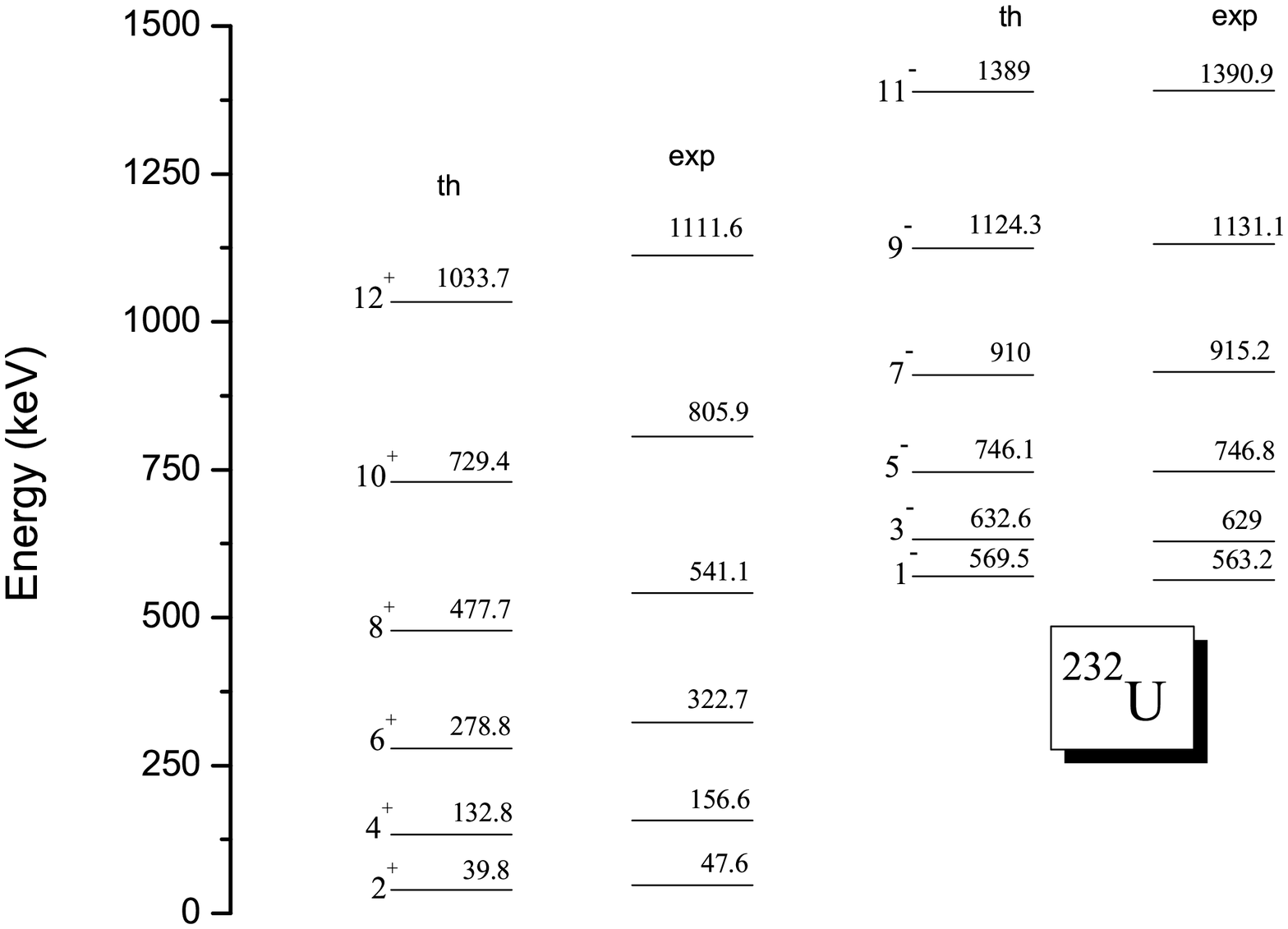}%
\includegraphics[width=0.5\textwidth]{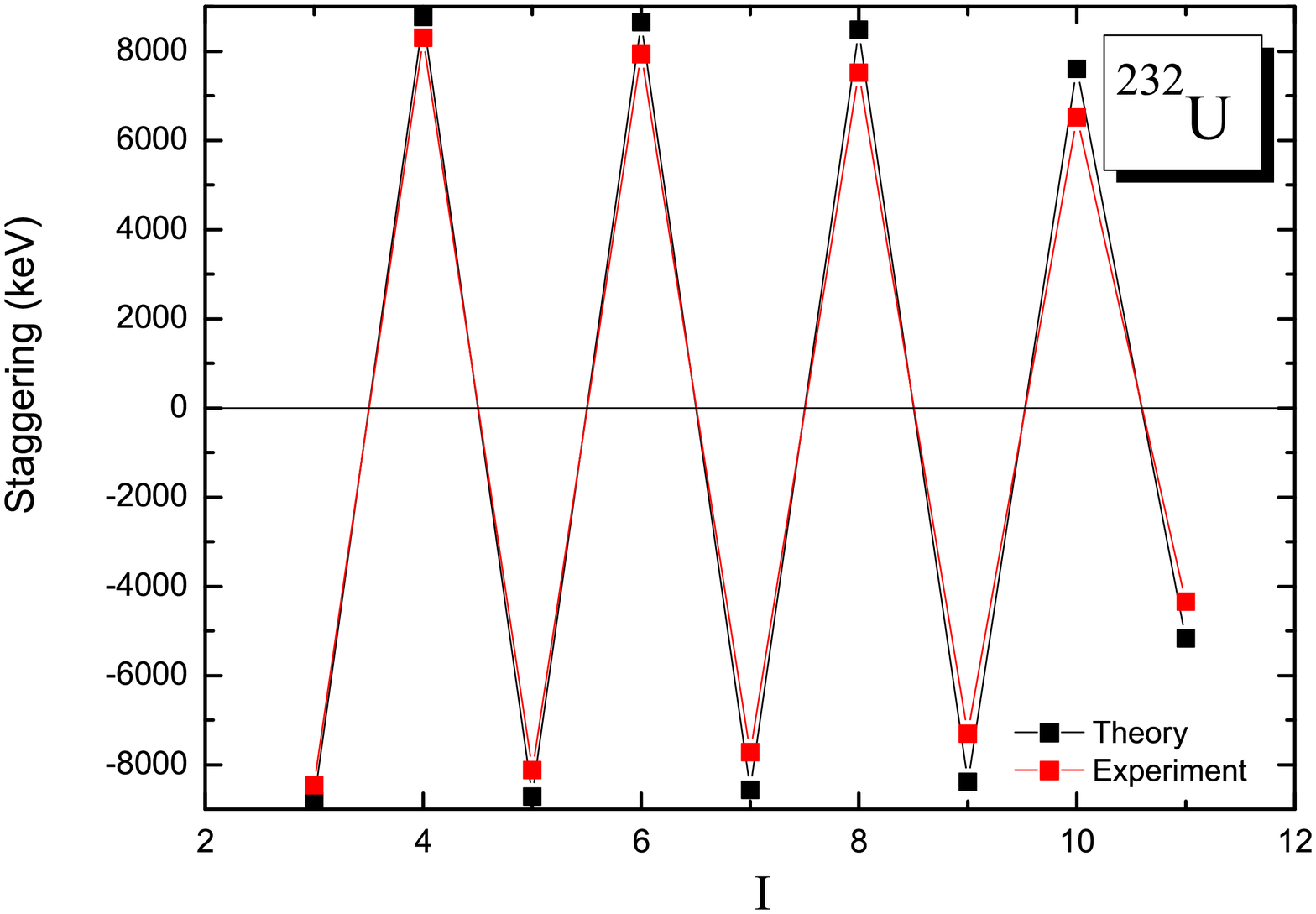}
\caption{The same as in Fig.1, but for $^{232}$U (parameters values:
 $\tilde{B}_2$=100$\hbar^{2}MeV^{-1}$,
$\tilde{B}_3$=1$\hbar^{2}MeV^{-1}$, $\gamma_{\mbox{\scriptsize eff}}$=55.4337$^{\circ}$,
$\eta_{\mbox{\scriptsize eff}}$=49.8625$^{\circ}$ and RMS=39.4 keV).}
\end{figure}

\begin{figure}
\includegraphics[width=0.5\textwidth]{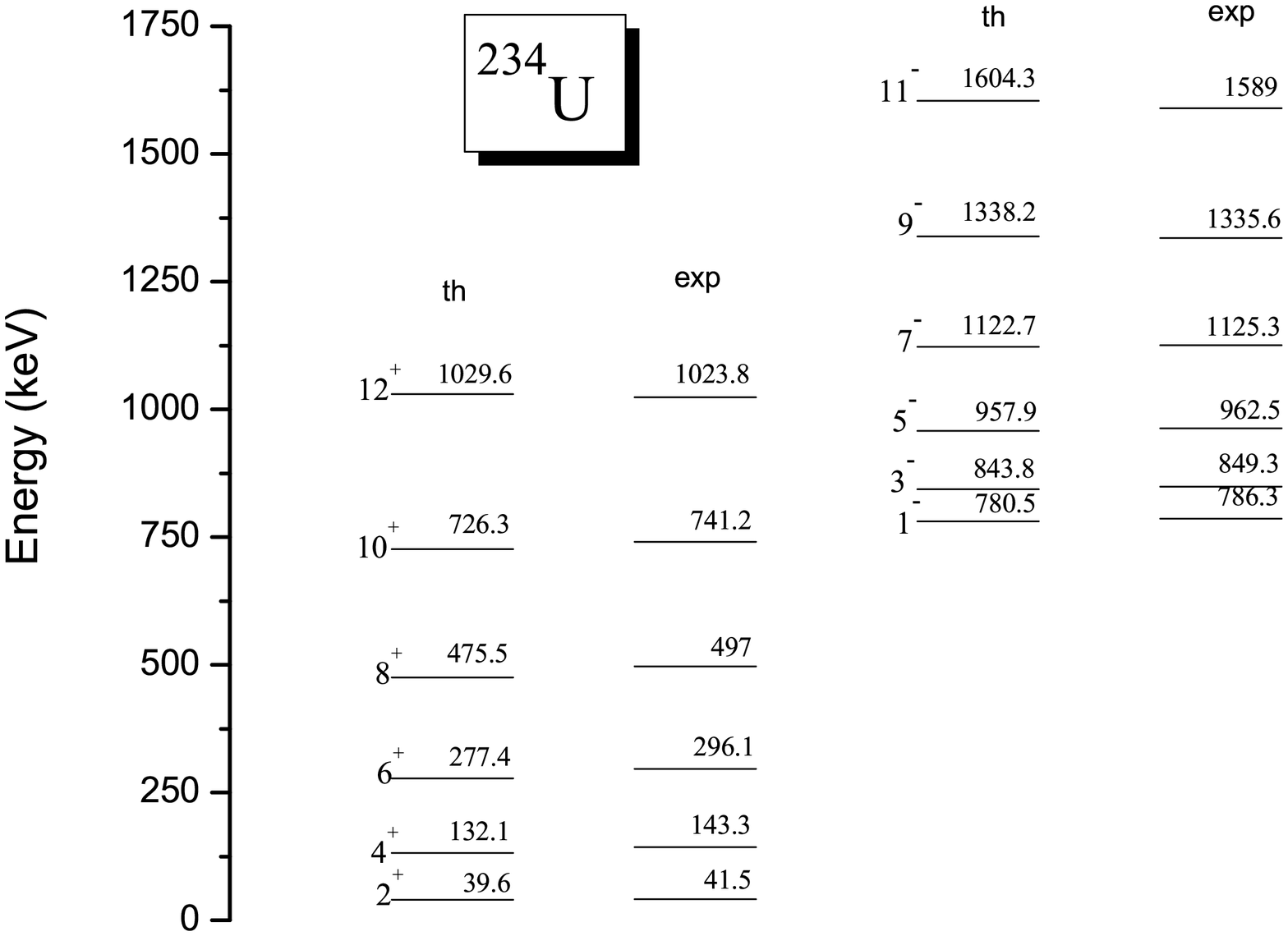}%
\includegraphics[width=0.5\textwidth]{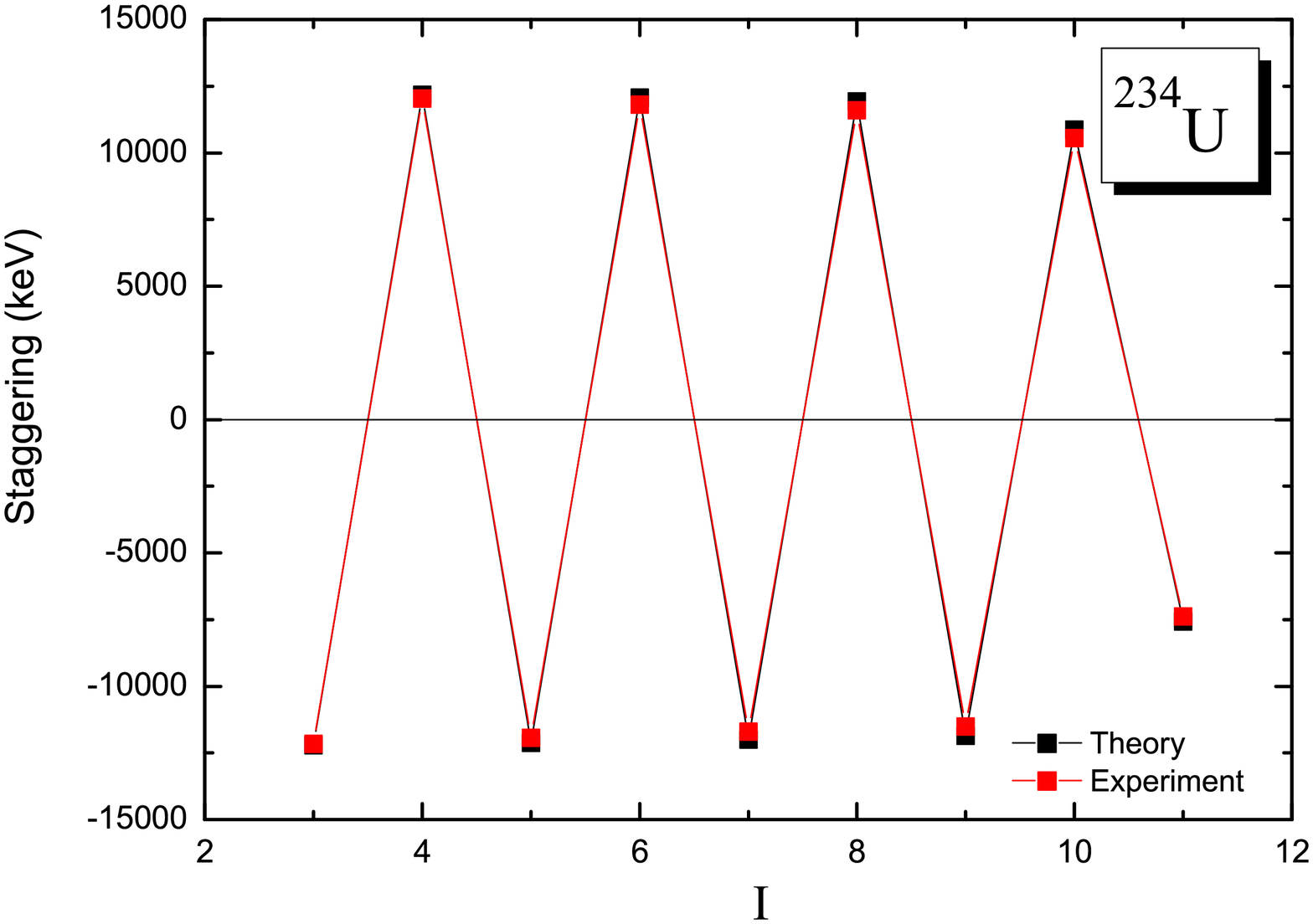}
\caption{The same as in Fig.1, but for $^{234}$U (parameters values:
 $\tilde{B}_2$=100$\hbar^{2}MeV^{-1}$,
$\tilde{B}_3$=1$\hbar^{2}MeV^{-1}$, $\gamma_{\mbox{\scriptsize eff}}$=56.4036$^{\circ}$,
$\eta_{\mbox{\scriptsize eff}}$=50.3398$^{\circ}$ and RMS=11.3 keV).}
\end{figure}

\begin{figure}
\includegraphics[width=0.5\textwidth]{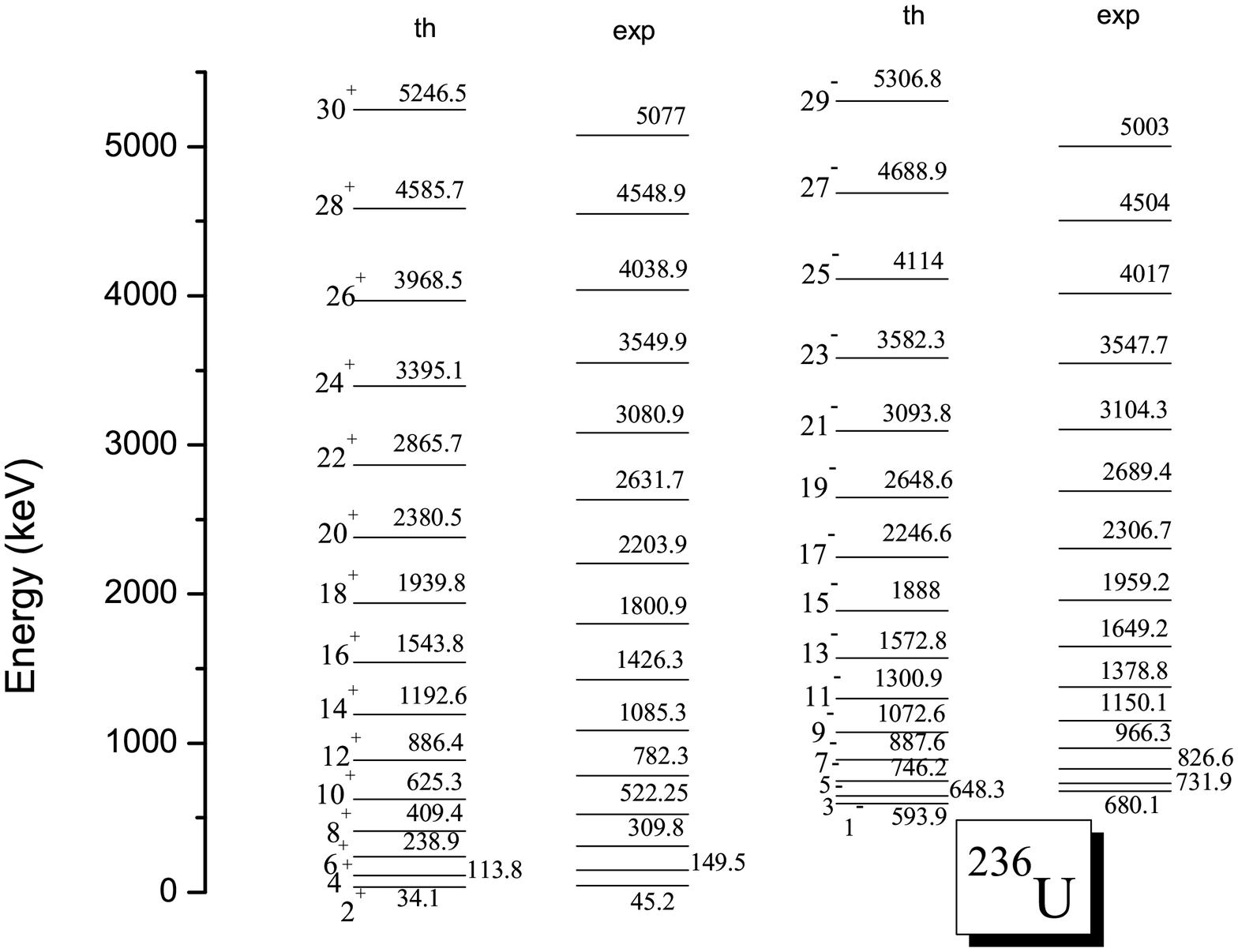}%
\includegraphics[width=0.5\textwidth]{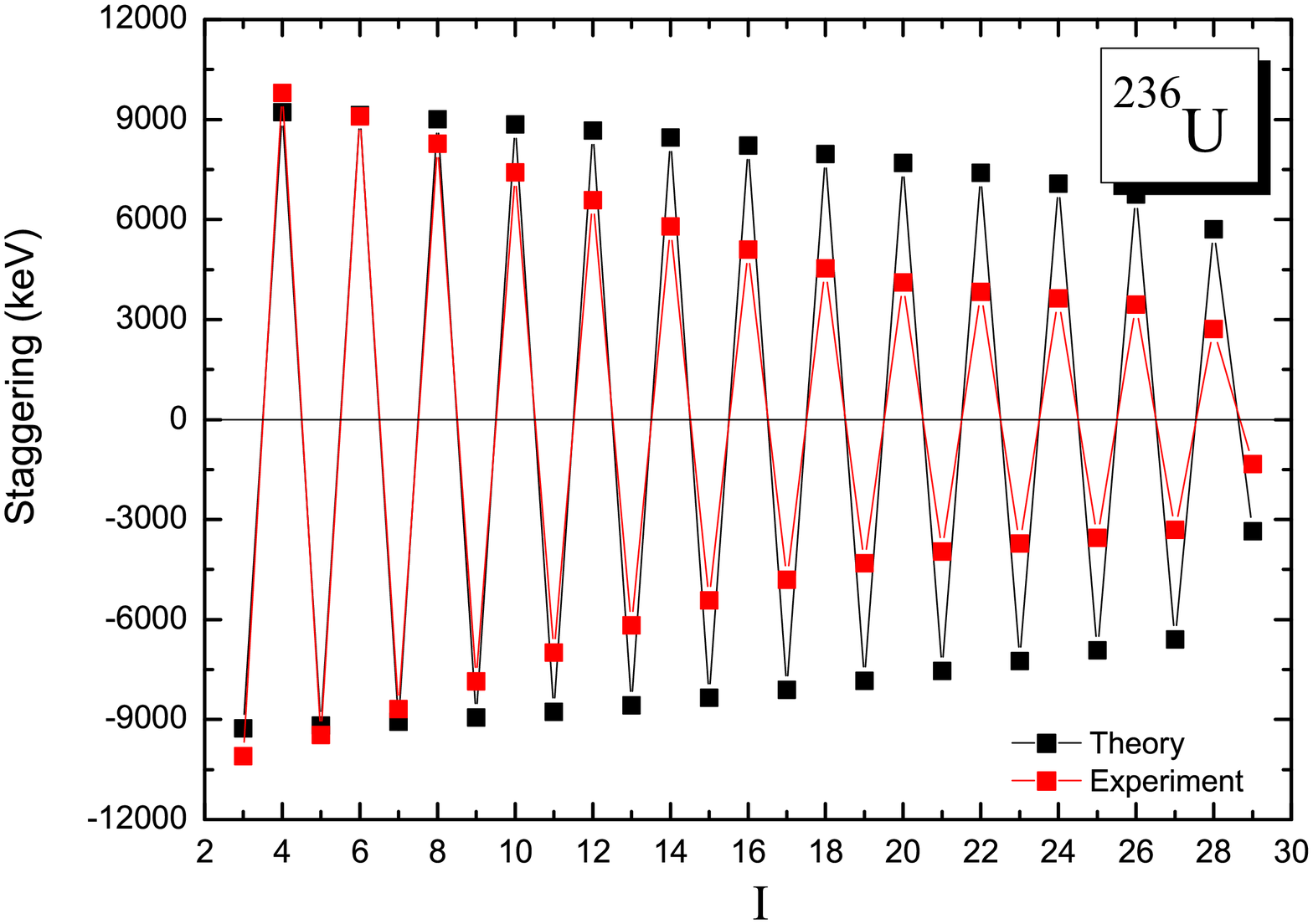}
\caption{The same as in Fig.1, but for $^{236}$U (parameters values:
 $\tilde{B}_2$=116.1$\hbar^{2}MeV^{-1}$,
$\tilde{B}_3$=1.268$\hbar^{2}MeV^{-1}$, $\gamma_{\mbox{\scriptsize eff}}$=56.1288$^{\circ}$,
$\eta_{\mbox{\scriptsize eff}}$=50.1802$^{\circ}$ and RMS=145.5 keV).}
\end{figure}

\begin{figure}
\includegraphics[width=0.5\textwidth]{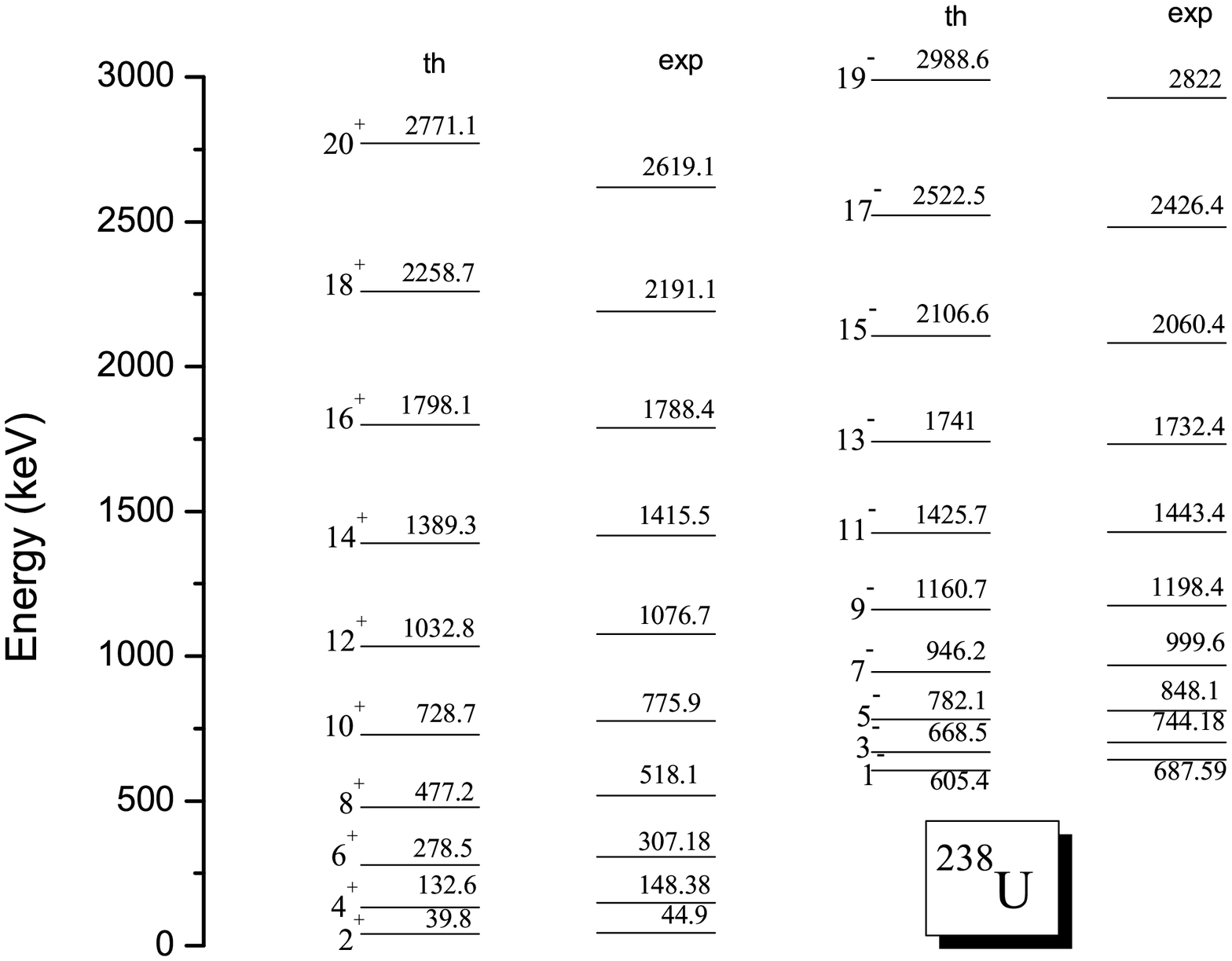}%
\includegraphics[width=0.5\textwidth]{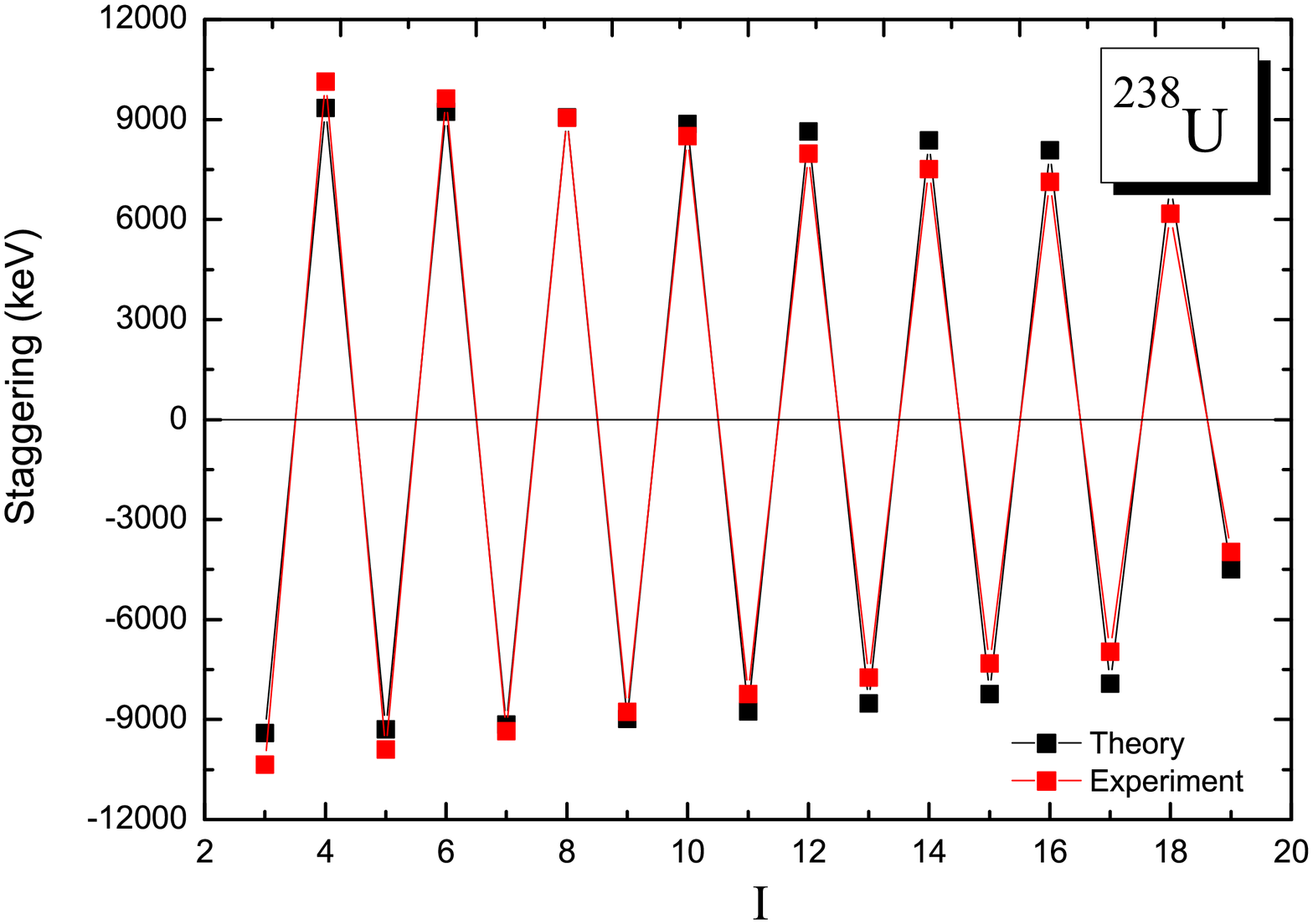}
\caption{The same as in Fig.1, but for $^{238}$U (parameters values:
 $\tilde{B}_2$=100$\hbar^{2}MeV^{-1}$,
$\tilde{B}_3$=1$\hbar^{2}MeV^{-1}$, $\gamma_{\mbox{\scriptsize eff}}$=55.6287$^{\circ}$,
$\eta_{\mbox{\scriptsize eff}}$=49.9292$^{\circ}$ and RMS=69.3 keV).}
\end{figure}

\begin{figure}
\includegraphics[width=0.5\textwidth]{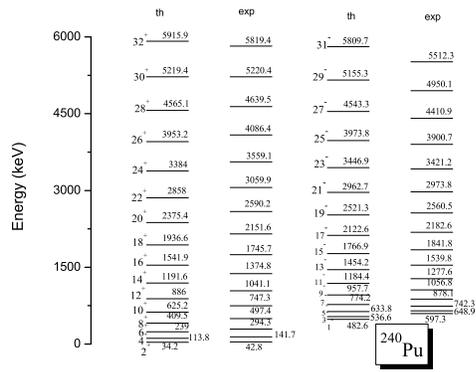}%
\includegraphics[width=0.5\textwidth]{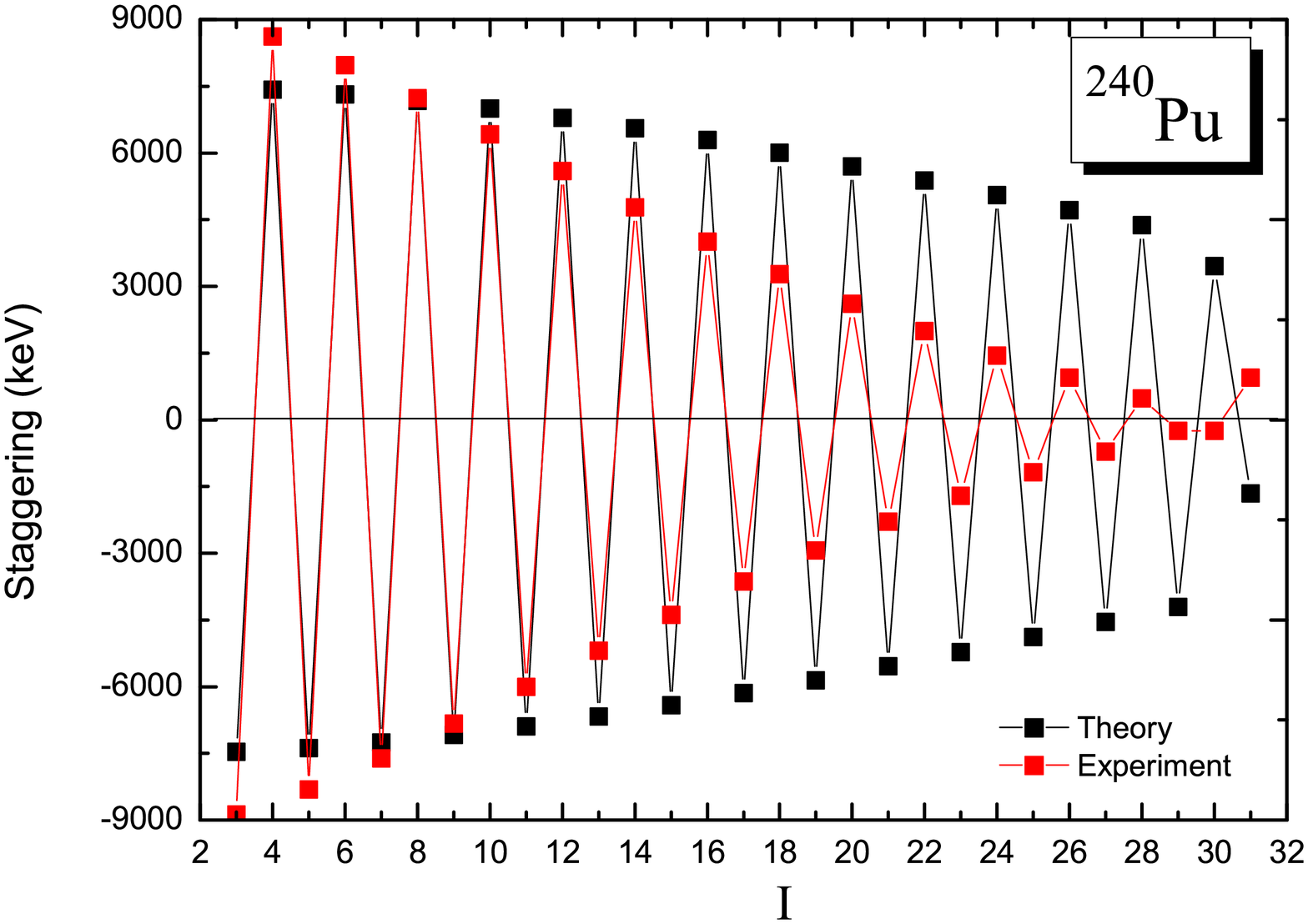}
\caption{The same as in Fig.1, but for $^{240}$Pu (parameters values:
 $\tilde{B}_2$=117$\hbar^{2}MeV^{-1}$,
$\tilde{B}_3$=1.02$\hbar^{2}MeV^{-1}$, $\gamma_{\mbox{\scriptsize eff}}$=55.2914$^{\circ}$,
$\eta_{\mbox{\scriptsize eff}}$=49.7773$^{\circ}$ and RMS=132.4 keV).}
\end{figure}
\end{document}